\documentclass[aps,twocolumn,prl]{revtex4}

\input epsf
\def\plotone#1{\centering \leavevmode
\epsfxsize= 1.0\columnwidth \epsfbox{#1}}

\def\apjl{Astrophys. J. Lett.}

\def\be{\begin{equation}}
\def\ee{\end{equation}}
\def\bea{\begin{eqnarray}}
\def\eea{\end{eqnarray}}

\def\GeV{\,{\rm GeV}}

\def\cmm2{{\,\rm cm^{-2}}}
\def\cm2{{\,{\rm cm}^2}}
\def\cmm3{{\,{\rm cm}^{-3}}}
\def\gcmm3{{\,{\rm g\,cm^{-3}}}}

\def\fun#1#2{\lower3.6pt\vbox{\baselineskip0pt\lineskip.9pt
  \ialign{$\mathsurround=0pt#1\hfil##\hfil$\crcr#2\crcr\sim\crcr}}}

\def\p3m{P$^3$M}

\def\ga{\mathrel{\mathpalette\fun >}}
\def\fun#1#2{\lower3.6pt\vbox{\baselineskip0pt\lineskip.9pt
  \ialign{$\mathsurround=0pt#1\hfil##\hfil$\crcr#2\crcr\sim\crcr}}}

\begin{document}
\title{A limit on the detectability of the energy scale of inflation
}
\author{Lloyd\ Knox and Yong-Seon\ Song}
\affiliation{Department of Physics, One Shields Avenue\\
University of California, Davis, California 95616, USA}
\date{\today}

\begin{abstract}
We show that the polarization of the cosmic microwave background can be used to
detect gravity waves from inflation if the energy scale of
inflation is above $3.2 \times 10^{15}$~GeV.  These gravity waves
generate polarization patterns with a curl, whereas (to first order
in perturbation theory) density perturbations 
do not.
The limiting ``noise'' arises from the second--order generation of curl from density
perturbations, or rather residuals from its subtraction.  We calculate
optimal sky coverage and detectability limits as a function of 
detector sensitivity and observing time. 
\end{abstract}
 \maketitle

Few ideas have had greater impact in cosmology than that of
inflation\cite{guth81,albrecht82,linde82}.  Inflation makes four
predictions, three of which provide very good descriptions of data:
the mean curvature of space is vanishingly close to zero, the power
spectrum of initial density perturbations is nearly scale--invariant,
and the perturbations follow a Gaussian distribution.  As the data
have improved substantially (e.g.,
\cite{halverson01,netterfield01,lee01}) they have agreed well with
inflation, whereas all competing models for explaining the
large--scale structure in the Universe have been ruled out (e.g.,
\cite{pen97,allen97,albrecht97}).

We must note though that these three predictions are all fairly
generic\footnote{For example, a much more impressive success would be
prediction of {\it features} in the power spectrum followed by their
detection.}.  Further, although existing models for
the formation of structure have been ruled out, there is no proof of
inflation's unique ability to lead to our Universe.  Indeed,
alternatives are being invented \cite{khoury01}.

The fourth (and yet untested) prediction may therefore play a
crucial role in distinguishing inflation from other possible early
Universe scenarios.  Inflation inevitably leads to a nearly
scale--invariant spectrum of gravitational waves, which are
tensor perturbations to the spatial metric.  Detection of
this gravitational--wave background might allow discrimination
between competing scenarios (e.g. \cite{khoury01}),  and
different inflationary models (e.g. \cite{dodelson97}).

The amplitude of the power spectrum of tensor perturbations to the
metric is directly
proportional to the energy scale of inflation.  One can
use a determination of the tensor contribution to CMB
temperature anisotropy, here parameterized by the quadruple
variance, to determine this energy scale \cite{turner95}:
\be
V_*^{1/4}/m_{Pl}   = 1.2\langle Q^2_T \rangle^{1/4} =
3.0 \times 10^{-3} r^{1/4}
\ee
where $r \equiv \langle Q^2_T\rangle/\langle Q^2_S\rangle$, $S$
stands for scalar (density) perturbation and $\langle Q^2_{S}\rangle \simeq
4 \times 10^{-11}$ from observations\footnote{The * subscript on
$V$ means that it is evaluated when the scale factor is about $e^{50}$
times smaller than it is at the end of inflation, when the relevant
perturbations for large--scale CMB fluctuations are exiting the
Horizon.  There is a slight dependence on $\Omega_\Lambda$
\cite{knox95}, here taken to be $\Omega_\Lambda = 0.65$.  Other
cosmological parameters assumed throughout are $h=0.65$, $\Omega_b = 0.05$,
$\Omega_m=0.35$ and $\delta_H = 4.2 \times 10^{-5}$.}.  Without
detection of gravitational waves, the energy scale of inflation
remains uncertain by at least 12 orders of magnitude.  Pinning down
this energy scale could be crucial to understanding how
inflation arises in a fundamental theory of physics.

\begin{figure}[htbp]
  \begin{center}
    \plotone{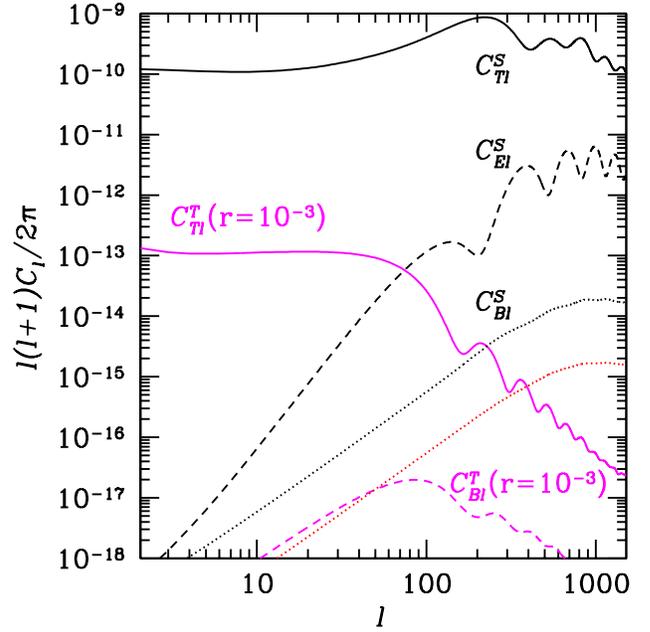}
    \caption{Angular power spectra. Solid lines are for
temperature anisotropies due to scalar perturbations, $C_{Tl}^S$
and tensor perturbations, $C_{Tl}^T$ with $r=10^{-3}$.  Dashed
lines are for the E modes from scalar perturbations, $C_{El}^S$
and the B modes from tensor perturbations, $C_{Bl}^T$.  The dotted
lines are for the lensing--induced scalar B-modes, $C_{Bl}^S$
before (above) and after (below) the cleaning that can be done by
a perfect experiment.
}
    \label{fig:fig1}
  \end{center}
\end{figure}

In Fig. 1 we show the angular power spectrum of CMB temperature
perturbations contributed by scalar perturbations and by tensor
perturbations with $r=10^{-3}$.  By determining the total CMB
temperature power spectrum we can determine or limit the energy
scale of inflation, based on the presence or absence of extra
power at low $l$.  The scalar temperature perturbations inevitably
limit our ability to detect the tensor temperature perturbations
to those cases with $r > r_{\rm lim} = 0.13$.
\cite{knox94}\footnote{Here and throughout, the detection limit is
set at 3.3 times the standard deviation.  For a normal
distribution this means that one can make a detection at 95\%
confidence for 95\% of possible realizations.}

In \cite{kamionkowski97, seljak97} it was pointed out that tensor
perturbations result in CMB polarization patterns with a curl,
whereas scalar perturbations do not.  By analogy with
electromagnetism, these modes are called ``B-modes'', and the
curl--free modes are called ``E-modes''.  This was an exciting
development, because the new signature of tensor perturbations
could not be confused with scalar perturbations. Figure 1 also
shows the power spectrum of this B-mode with the amplitude it
would have if $V_*^{1/4} = 6.4 \times 10^{15}$ GeV, corresponding
to $r=10^{-3}$.  It is a very weak signal, even at its peak
$~10^7$ times less than the power spectrum of the temperature
anisotropy. Ignoring any contaminating source of curl--mode
polarization the detectability limit is solely a matter of sufficient
sensitivity to the CMB polarization.  For a detector of
sensitivity $s$ uniformly observing the entire sky for time $t$:
\be \label{eqn:kamikows} r_{\rm lim} = 10^{-2} \left({s \over \mu
K \sqrt{\rm sec}}\right)^2\left({t \over 1 ~{\rm
year}}\right)^{-1}. \ee

Equation~\ref{eqn:kamikows} \cite{kamionkowski98} does not take
into account a contaminating source of B mode that arises from the
lensing of the E mode by density perturbations along
lines--of--sight between the observer and the last--scattering
surface \cite{zaldarriaga98}. This scalar contribution to the
B-mode power spectrum is shown in Fig. 1.  As we will see below
this contamination sets the detectability limit at 
$r_{\rm lim} =6 \times 10^{-4}$, similar to what was found in
\cite{lewis01}.

This lensing contaminant can be cleaned from the maps as discussed in
\cite{hu01, hu02}.  Here we show that there are limits to how well
these maps can be cleaned.  Our main result is that in the no--noise
limit the tensor perturbations can be detected only if $V_*^{1/4} >
3.2 \times 10^{15}\GeV$.  Interestingly, this is below the mass scale
for the gauge coupling--constant convergence at $10^{16}\GeV$ that
occurs in minimal extensions to the standard model of particle physics
(e.g.  \cite{langacker91}).

The Stokes parameters, $I$, $Q$ and $U$ are related to the
unlensed Stokes parameters (denoted with a tilde) by \bea
\label{eqn:deflect} I({\vec{\theta}}) = \tilde I( \vec{\theta}+ \delta
\vec{\theta}) & Q({\vec{\theta}}) = \tilde Q( \vec{\theta}+\delta
\vec{\theta}) & U({\vec{\theta}}) = \tilde U( \vec{\theta}+ \delta
\vec{\theta}). \eea The deflection angle, $\delta \vec{\theta}$, is the
tangential gradient of the projected gravitational potential, \be
\phi({\vec{\theta}})= 2\int dr{(r-r_s)\over r r_s}\Psi(r\hat n,r) \ee
where $r$ is the coordinate distance along our past light cone, $s$
denotes the CMB last--scattering surface and
$\hat n$ is the unit vector in the $\vec{\theta}$ direction.

The effect on the B--mode power spectrum is \cite{zaldarriaga98} 
\be C_{Bl} = C_{\tilde
B l} + \sum_{l'}W^{l'}_l C_{\tilde E l'} \ee where \bea W_l^{l'} =
{l'^3\over 4} \int_0^\pi \theta d\theta \times\{ \nonumber \\
 \sigma_2^2(\theta)
\left[J_0(l\theta)J_2(l'\theta)-{1 \over
2}J_4(l\theta)\left[J_2(l'\theta)+J_6(l'\theta)\right]\right]+
 \nonumber \\
\sigma_0^2(\theta)\left[J_4(l\theta)J_4(l'\theta)+
J_0(l'\theta)J_0(l\theta)\right]\}. \eea The functions
$\sigma_2^2(\theta)$ and $\sigma_0^2(\theta)$ depend on the
statistical properties of the displacement potential $\phi$ and
are given by \bea \sigma_0^2(\theta) &=& \int {ldl \over 2\pi} l^2
C_l^{\phi}(1-J_0(l\theta)) \nonumber \\
\sigma_2^2(\theta) &=& \int {ldl \over
2\pi}l^2C_l^{\phi}J_2(l\theta)\eea where \be C_l^{\phi} \equiv
\langle \phi_{lm}\phi_{lm}\rangle \ee and $\phi_{lm}$ is the
spherical--harmonic transform of $\phi({\vec{\theta}})$.

If we have a means of determining $\phi$, and therefore $\delta
\vec{\theta}$, we can reconstruct the unlensed maps from the lensed maps
by use of Eq.~\ref{eqn:deflect}.  This procedure cleans out the
lensing--induced B--mode. The $C_l^{\phi}$ above can either be
interpreted as the power spectrum of the lensing potential (in the
case of uncleaned maps in which case we will call it
$C_l^{\phi(S)}$) or the power spectrum of the lensing potential
residuals (in the case of lensing--cleaned maps).  In the
uncleaned case \cite{hu00} (in the Limber approximation valid at
small scales): \be \label{eqn:signal} C_l^{\phi(S)}
 = {8\pi^2 \over l^3}H_0^2\int_0^{r_s} dr r \left[{(r-r_s)\over rr_s}\right]^2\Delta_\Phi^2(k,r)|_{k=l/r} \ee
where $r_s$ is the comoving distance to the last-scattering surface
and $\Delta_\Phi^2(k,r) \equiv k^3/(2\pi)^2 P_\Phi(k,r)$ where
$P_{\Phi}(k,r)$ is the power spectrum of the gravitational potential
at the time corresponding to coordinate distance $r$ on our past light
cone.  We plot $C_l^{\phi(S)}$ in Fig. 2

We consider the $\phi$ reconstruction procedure given in
\cite{hu02} which exploits the fact that lensing leads to a
mode--mode coupling with expectation value proportional to $\phi$.
Their estimator is a minimum--variance (MV) average over pairs of map
modes with $l \ne l'$.  Since we know the statistics of the signal,
$C_l^\phi$, we can Wiener filter (WF) the MV estimate and further
reduce the errors in the reconstruction.  The error in the WF estimate
of $\phi_{lm}$ has variance:
\be
C^{\phi(WF)}_l = {C^{\phi(S)}_l C^{\phi(MV)}_l \over C^{\phi(S)}_l
+ C^{\phi(MV)}_l}. \ee The Wiener--filtering is important in the
low signal--to--noise regime $(C^{\phi(S)}_l/C^{\phi(MV)}_l \ll
1)$ where the variance of the reconstructed $\phi$ is
$C^{\phi(S)}_l$ instead of $C^{\phi(MV)}_l$.

\begin{figure}[htbp]
  \begin{center}
    \plotone{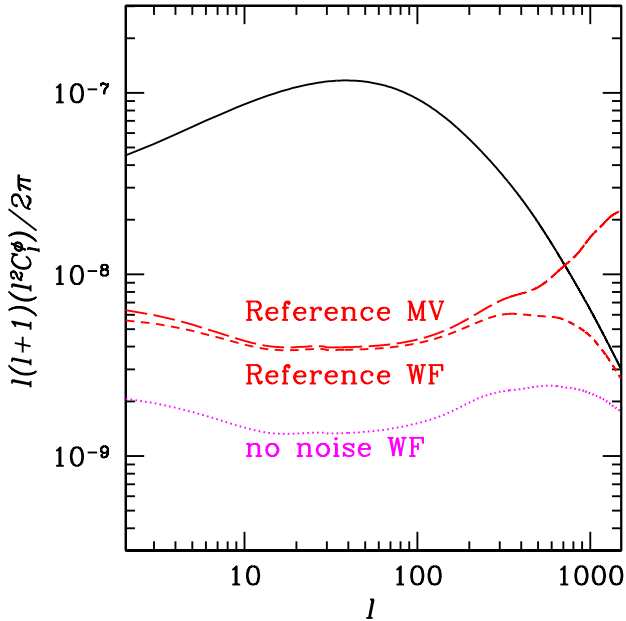}
    \caption{Angular power spectrum of projected gravitational
potential, $\phi$ (solid curve) and the power spectrum of the
residuals of different reconstruction procedures:  minimum variance
for ``reference'' experiment (long dashes), Wiener filter for the reference
experiment (short dashes) and Wiener filter for a noiseless experiment
(dotted line).  }
    \label{fig:fig2}
  \end{center}
\end{figure}

Cleaning can greatly reduce the amplitude of $C_l^\phi$ as is
shown in Fig. 2 where the angular power spectrum of the residual
$\phi$ is shown for two different experiments.  The first is the
`reference' experiment of \cite{hu02} which makes temperature maps
with weight-per-solid angle of $w = (1 \mu{\rm
K}-{\rm arcmin})^{-2}$ and $Q$ and $U$ maps each with half this
weight, all with $7'$ (full-width-half-maximum) resolution.  The
second is the no--noise, perfect angular resolution case.  One can
see that high signal--to--noise reconstruction of $\phi$ is
possible.

Because the cleaning is not perfect, there is still
lensing-induced scalar B-mode in the cleaned maps.  In the
no-noise limit the residual scalar B-mode power spectrum, $C_{Bl}^{WF}$, has
been reduced by a factor of 10 from the uncleaned amplitude, 
as shown in Fig. 1.  

Writing the tensor--induced B-mode power spectrum as $C_{Bl}^{T} =
rC_{Bl}^{T,1}$ the resulting error in $r$, $\sigma_r$, is given by \be
\label{eqn:fisher} {1 \over \sigma_r^2} =  f_{\rm sky}\sum_l {2l+1
\over 2}\left(C_{Bl}^{T,1}\right)^2
\left(C_{Bl}^{WF}+w^{-1}e^{l^2\sigma_b^2}\right)^{-2} \ee where
the $Q$ and $U$ maps cover $f_{\rm sky}$ of the sky with uniform
weight per solid angle $w$ and have been smoothed by a circular
Gaussian beam profile with full-width-half-maixmum = $\sqrt{8 \ln
2} \sigma_b$.  For $r_{\rm lim} = 3.3 \sigma_r$ we find for
noise--free maps, but with no cleaning done, $r_{\rm lim} = 6 \times 10^{-4}$.
For the `reference' experiment of \cite{hu01} we find $r_{\rm lim}
= 4 \times 10^{-4}$ after cleaning with the MV estimator and 
$1.8 \times 10^{-4}$ with the WF estimator.  Finally, for 
noise--free maps with WF cleaning $r_{\rm lim} = 6 \times 10^{-5}$.

All of the above results are for $f_{\rm sky} = 1$.  Given a
detector (or array of detectors) with total sensitivity $s$
and an amount of observing time $t$ there is an optimal amount
of sky to cover.  Too much sky coverage means the maps are noisy
and the lensing contaminant can not be removed accurately enough,
too little sky hampers the statistical subtraction of the residual
scalar B-mode power spectrum.  The optimal sky coverage is given
by $f_{\rm sky}^{\rm opt} = 10^{-8} (\mu {\rm K})^2 t/s^2$.  In Fig. 3 
we plot $r_{\rm lim}$ as a function of $s^2/t$ assuming optimal 
sky coverage (constrained to $f_{\rm sky} \le 1$).  The detectability
limit is not strongly sensitive to deviations from optimal sky coverage,
which is fortunate since other considerations can influence sky 
coverage choice as well.  Note that $s^2/t = 10^{-8} (\mu {\rm K})^2$
could be achieved by observing for a year with an array of 30,000
detectors each with sensitivity $100 \mu {\rm K} \sqrt{\rm sec}$. Ten
times more detectors (or integration time) are required to reach
$s^2/t = 10^{-9} (\mu {\rm K})^2$ and approach the smallest 
possible $r_{\rm lim}$.  We have ignored aliasing of the scalar
E--mode wich can be important for $f_{\rm sky} \ne 1$, as studied
in \cite{lewis01}. Aliasing considerations increase
$f_{\rm sky}^{\rm opt}$ and $r_{\rm lim}$.  
For $s^2/t = 10^{-7}(\mu {\rm K})^2$ 
and $f_{\rm sky}=1$, $r_{\rm lim} = 7\times 10^{-4}$. 
Therefore, including effects of aliasing would not increase $r_{\rm lim}$ 
by more than 40\% over our results as shown in Fig. 3.

\begin{figure}[htbp]
  \begin{center}
    \plotone{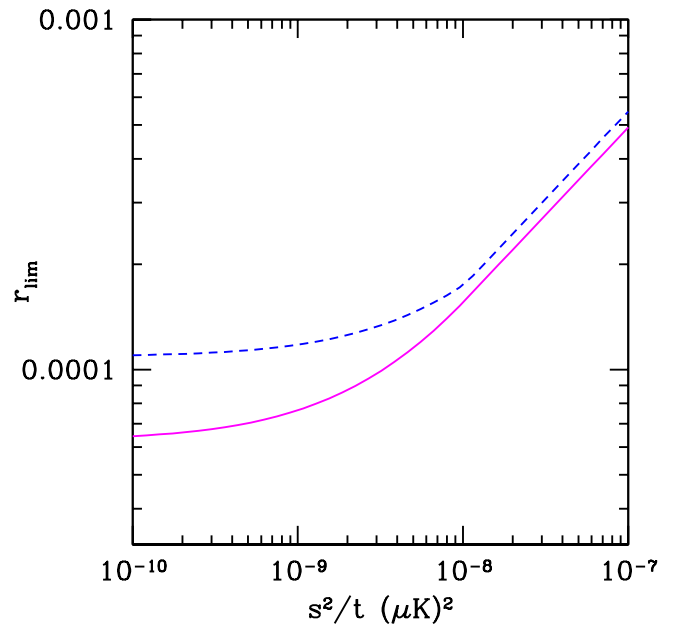}
    \caption{Achievable detectability limit as a function of
total detector array sensitivity, $s$ and observing time $t$, assuming
optimal sky coverage (lower curve) and 1/3 of optimal sky coverage 
(upper curve).  Detectability limit assuming three times optimal sky 
coverage overlaps the upper curve when optimal sky coverage 
is less than 1/3.
}
    \label{fig:fig3}
  \end{center}
\end{figure}

Since we are interested in determining a fundamental limit to the
detectability of tensor perturbations, undeterred by the virtual cost
of imaginary experiments, we explore one further method of cleaning
out the lensing contaminant.  Imagine we reconstruct the projected
gravitational potential or the density field on our past light cone
out to some limiting redshift, $z_{\rm lim}$\footnote{Perhaps this is
done with a POTENT--like\cite{bertschinger89} 
use of peculiar velocity measurements, or
analysis of the statistical distribution of apparent galaxy shapes.}.
Then we could directly calculate the lensing generated by the matter
at $z < z_{\rm lim}$.  Subtracting this lensing contaminant from CMB
maps leaves only the contribution from $z > z_{\rm lim}$.  The effect
of this remaining lensing can be calculated by altering the
lower limit of integration in Eq.~\ref{eqn:signal} from 
0 to $r(z_{\rm lim})$.  To improve upon the
noise--free self--cleaned maps result of $r_{\rm lim} = 6 \times
10^{-5}$ one needs to take $z_{\rm lim} \ga 7$.  We know of no way of
obtaining this information at such high $z$.

Our analysis ignores polarized emission from galactic and
extragalactic sources.  Multi--frequency
observations can be used to clean out these signals based on their
distinct spectral shapes.  However, even in the no--noise limit these
cannot be cleaned perfectly since their spectral shapes are not
perfectly well known and vary spatially (e.g. \cite{tegmark00b}).  
Our estimate of the detectability limit should be viewed as a lower limit.

How likely is it that the energy scale of inflation is high enough to
be determined?  
It is conceivable that $V_*^{1/4}$ is as
low as $10^3\GeV$\cite{knox93}.  Given even odds between $10^3$ and
$10^{19}$ GeV the chances are small.  On the other hand gauge coupling
unification gives us a hint that something interesting may be occurring
at $10^{16}$ GeV.  
If inflation has anything to do with grand
unification, or physics at higher energy scales, then the chances
are good.

The likelihood of detection may also be addressed by better
determination of the scalar spectrum.  If we assume particular
functional forms for $V(\phi)$ and a scalar perturbation spectrum with 
power--spectral index, $n$, near the scale--invariant
value of unity, then we can give constraints on the range of possible
values of $r$.  Here we follow the nomenclature and calculations of
\cite{dodelson97}.  Exponential potentials have 
$r = 5(1-n)$ and ``small--field polynomial'' potentials ($V(\phi) =
\Lambda^4\left[1-\left(\phi/\mu\right)^p\right]$) with $p=2$ have
$r = 10\exp\left(-50\left(1-n\right)\right)$.  Still other classes of
models (hybrid inflation and the small--field polynomial potentials
with $p > 2$) leave little or no relationship between $r$ and $n$;
$r$ can take on detectable values or vanishingly small ones.
Perhaps the happiest case is the simplest one:
polynomial potentials with $V(\phi)=\Lambda^4(\phi/\mu)^p$.  With $p
\ge 2$ all have $r > 0.1$, well above our detectability limit.

In sum, we have calculated how large the amplitude of tensor
perturbations must be in order for their effect on the CMB to be
distinguishable from those of lensing.  By reconstructing the lensing
potential, using the mode--mode coupling induced by lensing, one
can reduce the lensing contamination but not eliminate it.  The residual
lensing signal prevents detection of tensor perturbations if 
$V_* < (3.2 \times 10^{15}\GeV)^4$.  If $V_*$ is slightly larger then
an ambitious observational program may succeed in determining the
energy scale of inflation---possibly a key step towards
finding inflation a comfortable home in a fundamental theory of
physics.

We thank A. Albrecht, D. Chung, N. Dalal and J. Kiskis 
for useful conversations.  We used CMBfast \cite{seljak96}.

\end{document}